\documentclass[journal]{IEEEtran}

\ifCLASSINFOpdf
\else
   \usepackage[dvips]{graphicx}
\fi
\usepackage{url}

\usepackage{graphicx}
\usepackage{amssymb} 
\usepackage{epstopdf}
\usepackage[ruled,norelsize]{algorithm2e}
\makeatletter
\newcommand{\removelatexerror}{\let\@latex@error\@gobble}
\makeatother
\usepackage{pseudocode}
\usepackage{amsthm}
\usepackage{amsmath}
\usepackage{cite}
\usepackage{tikz}
\usepackage{marginnote}
\usepackage{multirow}
\usepackage{subfigure}
\usepackage{hyperref}
\usepackage{mathtools}
\usepackage{floatrow}
\usepackage{bm}
\makeatletter
\newcommand*\bigcdot{\mathpalette\bigcdot@{.65}}
\newcommand*\bigcdot@[2]{\mathbin{\vcenter{\hbox{\scalebox{#2}{$\m@th#1\bullet$}}}}}
\makeatother

\usepackage{booktabs,array}

\newcount\rowc

\makeatletter
\def\ttabular{%
	\hbox\bgroup
	\let\\\cr
	\def\rulea{\ifnum\rowc=\@ne \hrule height 1.3pt \fi}
	\def\ruleb{
		\ifnum\rowc=1\hrule height 1.3pt \else
		\ifnum\rowc=6\hrule height \heavyrulewidth 
		\else \hrule height \lightrulewidth\fi\fi}
	\valign\bgroup
	\global\rowc\@ne
	\rulea
	\hbox to 10em{\strut \hfill##\hfill}%
	\ruleb
	&&%
	\global\advance\rowc\@ne
	\hbox to 10em{\strut\hfill##\hfill}%
	\ruleb
	\cr}
\def\endttabular{%
	\crcr\egroup\egroup}

\begin{document}

\title{Variational manifold learning from incomplete data: application to multislice dynamic MRI}

\author{Qing Zou, Abdul Haseeb Ahmed, Prashant Nagpal, Sarv Priya, Rolf F Schulte, Mathews Jacob
\thanks{Qing Zou and Mathews Jacob are with the Department of Electrical and Computer Engineering, The University of Iowa, Iowa City, IA, USA (e-mail: zou-qing@uiowa.edu and mathews-jacob@uiowa.edu). Abdul Haseeb Ahmed is with Philips Healthcare, Rochester, MN, USA (e-mail: hasib\_bhati@yahoo.com). Prashant Nagpal is with the Department of Radiology, University of Wisconsin-Madison, Madison, WI, USA (email: pnagpal@wisc.edu). Sarv Priya is with the Department of Radiology, The University of Iowa, Iowa City, IA, USA (e-mail: sarv-priya@uiowa.edu). Rolf F Schulte is with General Electric Healthcare, Munich, Germany (email: schulte@ge.com). This work is supported by NIH under Grants R01EB019961 and R01AG067078-01A1. This work was conducted on an MRI instrument funded by 1S10OD025025-01.}}

\maketitle

\begin{abstract}
Current deep learning-based manifold learning algorithms such as the variational autoencoder (VAE) require fully sampled data to learn the probability density of real-world datasets. Once learned, the density can be used for a variety of tasks, including data imputation. However, fully sampled data is often unavailable in a variety of problems, including the recovery of dynamic and high-resolution MRI data considered in this work. To overcome this problem, we introduce a novel variational approach to learn a manifold from undersampled data. The VAE uses a decoder fed by latent vectors, drawn from a conditional density estimated from the fully sampled images using an encoder. Since fully sampled images are not available in our setting, we approximate the conditional density of the latent vectors by a parametric model whose parameters are estimated from the undersampled measurements using back-propagation. We use the framework for the joint alignment and recovery of multislice free breathing and ungated cardiac MRI data from highly undersampled measurements. Most of the current self-gating and manifold cardiac MRI approaches consider the independent recovery of images from each slice; these methods are not capable of exploiting the inter-slice redundancies in the datasets and require sophisticated post-processing or manual approaches to align the images from different slices. By contrast, the proposed scheme is able to align the multislice data and exploit the redundancies. Experimental results demonstrate the utility of the proposed scheme in dynamic imaging alignment and reconstructions.

\end{abstract}


\IEEEpeerreviewmaketitle

\section{Introduction}

\IEEEPARstart{D}{eep} generative models \cite{hinton2006reducing} that rely on convolutional neural networks (CNNs) are now widely used to represent data living on nonlinear manifolds. For instance, the variational autoencoder (VAE) \cite{kingma2013auto} represents the data points as CNN mappings of the latent vectors, whose parameters are learned using the maximum likelihood formulation. Since the exact log-likelihood of the data points is intractable, VAE relies on the maximization of a lower bound of the likelihood, involving an approximation for the conditional density of the latent variable represented by an encoder neural network. The VAE framework offers several benefits over the vanilla autoencoder \cite{hinton2006reducing}, including improved generalization \cite{higgins2016early} and ability to disentangle the important latent factors \cite{higgins2016beta,chung2015recurrent}. Unfortunately, most of the current generative models are learned from fully sampled datasets. Once learned, the probability density of the data can be used as a prior for various applications, including data imputation \cite{li2021separation,mani2021qmodel}. Unfortunately, fully-sampled datasets are often not available in many high-resolution structural and dynamic imaging applications to train autoencoder networks. 

The main focus of this paper is to introduce a variational framework to learn a deep generative manifold directly from undersampled/incomplete measurements. The main application motivating this work is the multislice free-breathing and ungated cardiac MRI. Breath-held CINE imaging, which provides valuable indicators of abnormal structure and function, is an integral part of cardiac MRI exams. 
Compressed sensing \cite{kido2016compressed,axel2016accelerated,lustig2007sparse,Kustner2020} and deep learning methods have emerged as powerful options to reduce the breath-hold duration, with excellent performance \cite{Qin2019,prietoreview,kustner2020cinenet,sandino2021accelerating,wang2020ica}. Despite these advances, breath-held CINE imaging is challenging for several subject groups, including pediatric and chronic obstructive pulmonary disease (COPD) subjects. Several authors have introduced self-gating \cite{christodoulou2018magnetic,feng2016xd,feng2014golden,deng2016four,rosenzweig2020cardiac,Zhou2019} and manifold approaches \cite{Usman2015,Chen2017,nakarmi2018mls,shetty2019bi,nakarmi2017kernel} to enable free-breathing and ungated single-slice cardiac MRI. For instance, the smoothness regularization on manifolds (SToRM) approach \cite{ahmed2020free,poddar2015dynamic,poddar2019manifold} models the images as points on a low-dimensional manifold whose structure is exploited using a kernel low-rank formulation \cite{poddar2015dynamic,poddar2019manifold} to recover the images from highly undersampled measurements. Recently, deep learning-based manifold models were introduced \cite{zou2021dynamic,zou2021deep,yoo2021time} to further improve the performance; these schemes learn a deep generative network and its latent variables directly from the measured k-space data using a non-probabilistic formulation. 

All of the previously described free-breathing cardiac MRI reconstruction approaches (e.g., compressed sensing-based approaches, manifold approaches, and deep learning-based approaches) independently recover the data from each slice. Cardiac MRI often relies on slice-by-slice acquisition to preserve myocardium to blood pool contrast, resulting from the in-flow of blood from unexcited regions to the slice of interest; the improved contrast facilitates segmentation.  The above-mentioned 2D self-gating and manifold methods are thus unable to exploit the extensive redundancies between adjacent slices, which could offer improved performance. Note that the respiratory and cardiac motion during the acquisition of the different slices is often very different; this makes the direct 3D extension of the 2D self-gating and manifold methods impossible. Another challenge with the approaches mentioned above is the need for post-processing methods to determine matching slices at specific cardiac/respiratory phases for estimation of cardiac parameters (e.g., ejection fraction, strain). Several post-processing methods have been introduced to align the data post reconstruction \cite{Chen2017,gori2005exact,baumgartner2014high,baumgartner2015self,chen2016dynamic}. Because these methods require fully sampled data, they will not facilitate the exploitation of the inter-slice redundancies during image recovery. 

We introduce a novel variational framework for the joint recovery and alignment of multislice data from the entire heart. This approach combines the undersampled k-t space data from different slices, possibly acquired with multiple cardiac and respiratory motion patterns, to recover the 3D dynamic MRI dataset. We use a 3D CNN generative model, which takes in a latent vector and outputs a 3D image volume.  The time-varying latent vectors capture the intrinsic variability in the dataset, including cardiac and respiratory motion. The latent variables and the parameters of the 3D CNN are jointly learned from the multislice k-t space data using a maximum likelihood formulation. Since the likelihood is not tractable, we maximize its variational lower bound involving a model for the conditional distribution of the latent variables, which is conceptually similar to the VAE approach \cite{kingma2013auto}. The VAE scheme uses an encoder network to derive the conditional probabilities of the latent vectors from fully sampled data \cite{kingma2013auto}. This approach is not directly applicable in our setting because each data sample is measured using a different measurement operator. We hence model the conditional densities as a Gaussian distribution whose parameters are learned from the undersampled data directly using back-propagation. We use a Gaussian prior on the latent variables while deriving the evidence-based lower bound (ELBO); the Gaussian prior ensures that the latent variables from different slices have similar distributions, facilitating the alignment of the slices. We note that the direct extension of our previous generative manifold model \cite{zou2021dynamic,zou2021deep} to the 3D setting does not have any constraint on the latent variables; this extension results in poor alignment of the slices and degradation in image quality in the 3D setting. We also use smoothness priors on the latent variables to further improve the performance. Once learned, the representation can be used to generate matching 3D volumes with any desired cardiac/respiratory phase by exciting the generator with appropriate latent vectors. This approach of learning a generative model of the entire heart may thus be viewed as a paradigm shift from conventional slice-by-slice image-recovery algorithms\cite{christodoulou2018magnetic,feng2016xd,feng2014golden,deng2016four,rosenzweig2020cardiac,Zhou2019,Usman2015,Chen2017,nakarmi2018mls,shetty2019bi,nakarmi2017kernel,ahmed2020free,poddar2015dynamic,poddar2019manifold}.

\section{Background on dynamic MRI}

\subsection{multislice free-breathing MRI: problem statement}
The main application considered in this paper is the recovery of 3D cardiac volumes of the heart from undersampled 2D multislice k-t space data acquired in the free-breathing and ungated setting. In particular, we consider the recovery of the time series $\mathbf{x}(\mathbf{r},t_z)$, where $\mathbf r=(x,y,z)$ represents the spatial coordinates and $t_z$ denotes the  time frame during the acquisition of the $z^{\rm th}$ slice. We model the acquisition of the data as 
\begin{equation}
\mathbf b(t_z) = \mathcal{A}_{t_z}\Big(\mathbf{x}(\mathbf r,t_z)\Big) + \mathbf n_{t_z},
\end{equation}
where $\mathbf b(t_z) $ is the k-t space data of the $z^{\rm th}$ slice at the $t^{\rm th}$ time frame. Here, $\mathcal{A}_{t_z}$ are the time-dependent measurement operators, which evaluate the multi-channel single-slice Fourier measurements of the 3D volume $\mathbf x(\mathbf r,t_z)$ on the trajectory $k_{t_z}$ corresponding to the time point $t$. Specifically, $\mathcal{A}_{t_z}$ extracts the $z^{\rm th}$ slice from the volume $\mathbf x(\mathbf r, t_z)$ and evaluates its single-slice measurements. $\mathbf n_{t_z}$ represents the noise in the measurements.

\subsection{CNN-based generative manifold models in dynamic MRI}

CNN-based generative models were recently introduced for single-slice dynamic MRI \cite{zou2021dynamic}. This scheme models the 2-D images in the time series as the output of a CNN generator $\mathcal{D}_{\theta}$:
\[\mathbf{x}_i = \mathcal{D}_{\theta}(\mathbf{c}_i),\quad i = 1,\cdots, M.\]
The input $\mathbf{c}_i$ is the latent vector, which lives in a low-dimensional subspace. The recovery of the images in the time series involves the minimization of the criterion
\begin{align}\label{gen_SToRM}\nonumber
\mathcal C(\mathbf c,\theta)=& \underbrace{\sum_{i=1}^N\|\mathcal A_i\left(\mathcal D_{\theta}(\mathbf c_i)\right) - \mathbf b_i\|^2}_{\scriptsize\mbox{data term}} \\ 
&+ \lambda_1 \underbrace{\|J_{\mathbf c} \mathcal D_{\theta}(\mathbf{c})\|^2}_{\scriptsize \mbox{net reg.}} +  \lambda_2 \underbrace{\|\nabla_{i} \mathbf c_i\|^2 }_{\scriptsize\mbox{latent reg.}}.
\end{align}
The first term in the cost function is a measure of data consistency, while the second term is a network regularization term that controls the smoothness of the generated manifold \cite{zou2021dynamic}. The last term is the temporal smoothness of the latent variables, which is used to further improve the performance.

\section{Variational manifold learning}

We now introduce a novel variational formulation to learn a manifold from undersampled measurements, which is the generalization of the seminal VAE approach \cite{kingma2013auto} to the undersampled setting. We will first present the proposed approach in a simple and general setting for simplicity and ease of understanding. The use of this variational manifold model for the joint alignment and recovery of 3D images from 2-D multislice MRI data will be described in Section \ref{dmri}. 

\subsection{General problem statement and intuition}

We assume that the images in the time series, indexed by $i$, live on a smooth manifold $\mathcal M$ and hence can be modeled as the output of a CNN-based generator:
\begin{equation}\label{key}
\mathbf x_i = \mathcal D_{\theta}(\mathbf c_i),
\end{equation}
where $\mathbf c_i$ is the low-dimensional latent variable corresponding to $\mathbf x_i$. Here, $\theta$ denotes the weights of the generator, which is shared for all the images. 

Most generative models consider the learning of the above model from fully sampled data. By contrast, we consider the recovery from incomplete measurements 
\begin{equation}
\label{measurements}
\mathbf b_i = \mathcal A_i(\mathbf x_i) + \mathbf n_i.
\end{equation}
Here, $\mathcal A_i$ is an undersampled measurement operator corresponding to the $i^{\rm th}$ image frame. 
 Here, $\mathbf{n}_i\in\mathcal{N}(\mathbf{0},\sigma^2\mathbf{I})$ are noise vectors. Note that the measurement operators for each $\mathbf x_i$ are different. If the same sampling operators are used for all the data points, it is impossible to recover the images without additional prior information. We assume that the sampling operators satisfy the following properties:
\begin{enumerate}
	\item We assume  $\mathcal A_i$ to be a rectangular sub-matrix, obtained by picking specific rows of an orthonormal measurement operator (e.g., Fourier transform).
	\item We assume that the measurement operators $\mathcal A\sim \mathcal S$ are drawn from a distribution and satisfy
	\begin{equation}\label{expect}
	\mathbb E_{\mathcal A\sim S}[\mathcal A^T \mathcal A ] = \mathcal I,
	\end{equation}
	which is the identity operator. The above condition guarantees diversity in the measurement operators.
\end{enumerate}

We now provide some intuition about why the learning of the model with the above settings will succeed under the restrictive assumptions on the measurement operators described above. In the noiseless setting, we consider the learning of the latent variables $\mathbf c_i$ and the weights $\theta$ by minimizing the empirical error:

\begin{equation}\label{key}
\left\{ \theta^*,\mathbf c_i^*\right\} = \arg \min_{ \theta,\mathbf c_i} \underbrace{\sum_i \|\mathcal A_i\left(\mathbf x_i -  \mathcal D_{\theta}(\mathbf c_i)\right)\|^2}_{\mathcal L}.
\end{equation}
Here, $\mathbf x_i$ are the fully sampled data points. When  $\mathcal A\sim \mathcal S$, this empirical sum approximates
\begin{eqnarray*}
\mathcal L &\approx&  ~\mathbb E_{\mathbf x \sim \mathcal{M}}~\mathbb E_{\mathcal A \sim S} \|\mathcal A\left(\mathbf x -  \mathcal D_{\theta}(\mathbf c)\right)\|^2\\ 
&=& \mathbb E_{\mathbf x \sim \mathcal{M}}~\mathbb E_{\mathcal A \sim S} \left\langle \mathbf x -  \mathcal D_{\theta}(\mathbf c),\mathcal A^H\mathcal A\left(\mathbf x -  \mathcal D_{\theta}(\mathbf c)\right)\right\rangle\\
&=& \mathbb E_{\mathbf x \sim \mathcal{M}}~ \left\langle \mathbf x -  \mathcal D_{\theta}(\mathbf c),\underbrace{\mathbb E_{\mathcal A \sim S}[\mathcal A^H\mathcal A]}_{\mathcal I}\left(\mathbf x -  \mathcal D_{\theta}(\mathbf c)\right)\right\rangle\\
 &=&  \arg \min_{ \theta,\mathbf c}\mathbb E_{\mathbf x \sim \mathcal M} ~\|\mathbf x -  \mathcal D_{\theta}(\mathbf c)\|^2.
\end{eqnarray*}
The above result follows from \eqref{expect} and the orthonormality of the full measurement operator. This result shows that the recovery of the true manifold is feasible from undersampled data when the sampling operators satisfy the properties listed above.

\subsection{Proposed algorithm}

We consider the recovery of the images $\mathbf x_i$ from their measurements \eqref{measurements} by maximizing their likelihood, specified by 
\begin{equation}\label{likelihood}
p(\mathbf{b}_i)= \frac{p(\mathbf{b}_i,\mathbf{c}_i)}{p(\mathbf{c}_i|\mathbf{b}_i)}
\end{equation}
We note that the posterior $p(\mathbf{c}_i|\mathbf{b}_i)$ is not tractable. Following the VAE approach in \cite{kingma2013auto}, we use a surrogate distribution to approximate $p(\mathbf{c}_i|\mathbf{b}_i)$. The VAE formulation uses an encoder network to model $p(\mathbf{c}_i|\mathbf{x}_i)$ from 
the fully sampled data ($\mathbf b_i=\mathbf x_i$). Unfortunately, this approach is not directly applicable in our setting since $\mathbf b_i$ is the undersampled data, measured using $\mathcal A_i$ that vary with $i$.  

We propose to use a Gaussian model $q_i(\mathbf{c}_i) \approx p(\mathbf{c}_i|\mathbf{b}_i)$, parameterized by its mean $\boldsymbol\mu_i$ and diagonal covariance matrix $\boldsymbol\Sigma_i$, and to estimate these parameters using back-propagation. Following a similar argument as in \cite{kingma2013auto}, we show in the Appendix that the likelihood term in \eqref{likelihood} can be lower-bounded as 
\begin{eqnarray}\label{elbo}\nonumber
\log p(\mathbf{b}_i) &\ge& \underbrace{-\frac{1}{2\sigma^2}\mathbb E_{\mathbf c_i \sim q_i(\mathbf c_i)}\left[ \|\mathcal{A}_i \,\mathcal D_{\theta}(\mathbf{c}_i)-\mathbf b_i\|^2\right]}_{{\text{data term}}} \\
&&\qquad -\qquad \underbrace{KL[q_i(\mathbf c_i)||p(\mathbf{c}_i)]}_{L( q_i):\text{latent regularization}}.
\end{eqnarray}
Here, $p(\mathbf c_i)$ is a prior on the latent variables. In this work, we assume $p(\mathbf c_i) = \mathcal{N}(\mathbf{0},\mathbf{I})$, where $\mathbf{I}$ is the identity matrix. In this case, the KL divergence can be explicitly evaluated as
\[L(\mathbf c_i) = \frac{-\log[\det(\mathbf{\Sigma})] - n + {\mathrm{trace}}(\mathbf{\Sigma}) + \bm{\mu}^T\bm{\mu}}2,\] 
where we assume a latent space of dimension $n$. 

We hence solve for the unknown weights of the generator $\theta$ as well as the parameters of $q_i$ denoted by  $\boldsymbol\mu_i$ and $\boldsymbol\Sigma_i$ by minimizing the negative of the lower bound in \eqref{elbo}.

 Following \cite{kingma2013auto}, we use a Monte-Carlo approach to approximate the expectation in the data term. In particular, at each epoch of the training loop, we derive the samples $\mathbf c_i$ as
\begin{equation}\label{resampling}
\mathbf c_i = \boldsymbol \mu_i + \boldsymbol \Sigma_i ~\boldsymbol\epsilon,
\end{equation}
where $\boldsymbol \epsilon$ is a zero-mean unit variance Gaussian random variable. At each iteration, the estimation process thus involves the minimization of the criterion
\begin{equation}\label{vloss}
\mathcal{C}\left(\theta,\{\underbrace{\boldsymbol\mu_i, \boldsymbol \Sigma_i}_{q_i}\}\right) = \sum_{i=1}^{N_{\rm data}} \left(\|\mathcal{A}_i \,\mathcal D_{\theta}(\mathbf{c}_i)-\mathbf b_i\|^2 +  \sigma^2~ L(q_i) \right),
\end{equation}
with respect to the unknowns $\theta,\boldsymbol\mu_i$ and $\boldsymbol \Sigma_i$. 

\begin{figure*}[!htpb]
	\centering
\includegraphics[width=0.6\textwidth]{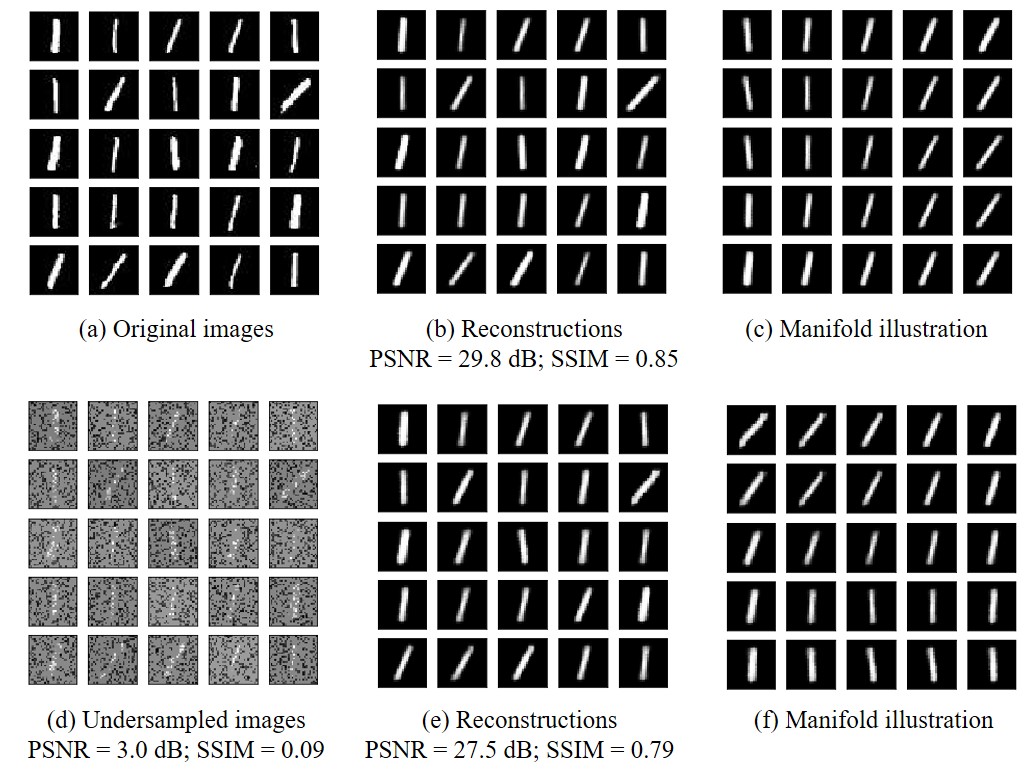}
	\caption{Illustration of variational manifold learning in the context of learning the digit 1 from the MNIST dataset. We first trained the variational model from the fully sampled data. (a) shows several of the original images, and (b) shows the corresponding output of the generator (reconstructions). (c) illustrates the learned manifold; we sample the latent vectors on a uniform grid in the range $[-3,3]^2$ and show the corresponding reconstructions. Note that the latent vectors capture the intrinsic variability in the dataset. The second row shows the results from the variational model, which are trained with undersampled noisy measurements. In this setting, 70\% of the pixel values are missing, and the remaining 30\% of the pixel values are corrupted with Gaussian white noise with 0 mean and 0.05 standard deviation. The zero-filled images are shown in (d). In (e), we show the reconstructions from the undersampled measurements. 
	Note that the reconstructions closely resemble the original digits in (a). (f) is the illustration of the manifold. Note that the variability in the manifold is captured in comparison to (c).  } 
	\label{mnist_example}
\end{figure*}

\subsection{Illustration using MNIST data}

We provide a simple example for the illustration of the above variational model from undersampled data of the digit 1 in the MNIST dataset \cite{lecun1998mnist}. The images used are scaled to the range $[-1,1]$.

The generator we used here is a simple CNN with three layers. ReLU activation function is used for the first two layers and tanh is used for the last layer. The dimension of the latent space is chosen as 2. In this example, all the trainable parameters are initialized as small random numbers, and the hyper-parameter for the latent regularization $L(\mathbf{c}_i)$ is chosen as 1. We used 1,000 epoches to train the CNN generator.

We first trained the model from the fully sampled data ($\mathcal A_i = \mathcal I$), whose results are shown in the first row of Fig. \ref{mnist_example}. Then we trained the model from undersampled noisy data. In the example, 70\% of the pixel values in each image are missing, while Gaussian white noise with standard deviation 0.05 is added to the known 30\% pixel values. The recovered images are shown in the second row of Fig. \ref{mnist_example}. We report the peak signal-to-noise ratio (PSNR) and the structural similarity index measure (SSIM) for the results.

\section{Application to dynamic MRI}
\label{dmri}
We first describe the application of the algorithm in the single-slice free-breathing and ungated data, which is the setting considered in \cite{zou2021dynamic}. 
We then generalize the approach to the novel setting of the joint alignment and recovery of 3D MRI from multislice free-breathing data in Section \ref{multislice}. 

\subsection{Acquisition scheme and pre-processing of data}

The datasets used in this work are acquired using a 2D (GRE) sequence with golden angle spiral readouts in the free-breathing and ungated setting on a MR750W scanner (GE Healthcare, Waukesha, WI, USA). The sequence parameters for the datasets are: FOV = 320 mm $\times$ 320 mm, flip angle = 18$^{\circ}$, slice thickness = 8 mm. The datasets were acquired using a cardiac multi-channel array with 34 channels. The Institutional Review Board at the University of Iowa approved the acquisition of the data, and written consents were obtained from the subjects. The number of slices acquired for different subjects varies. 

We used an algorithm developed in house to pre-select the coils that provide the best signal-to-noise ratio in the region of interest.  A PCA-based coil combination scheme was then used such that the approximation error was less than $5\%$. We then estimated the coil sensitivity maps based on these virtual channels using ESPIRiT \cite{uecker2014espirit} and assumed them to be constant over time. 

A total of 3,192 spirals were acquired for each slice in the subjects with TR=8.4 ms, which corresponds to an acquisition time of 27 seconds. Among the 3,192 spirals, every sixth spiral was acquired with the same angle; these spirals were used for self-navigation in the reconstruction methods that require self-navigation. We binned the data from six spiral interleaves corresponding to 50 ms temporal resolution for each frame.

\begin{figure*}[!htpb]
\centering
           \subfigure[V-SToRM: SS]{\includegraphics[width=0.38\textwidth]{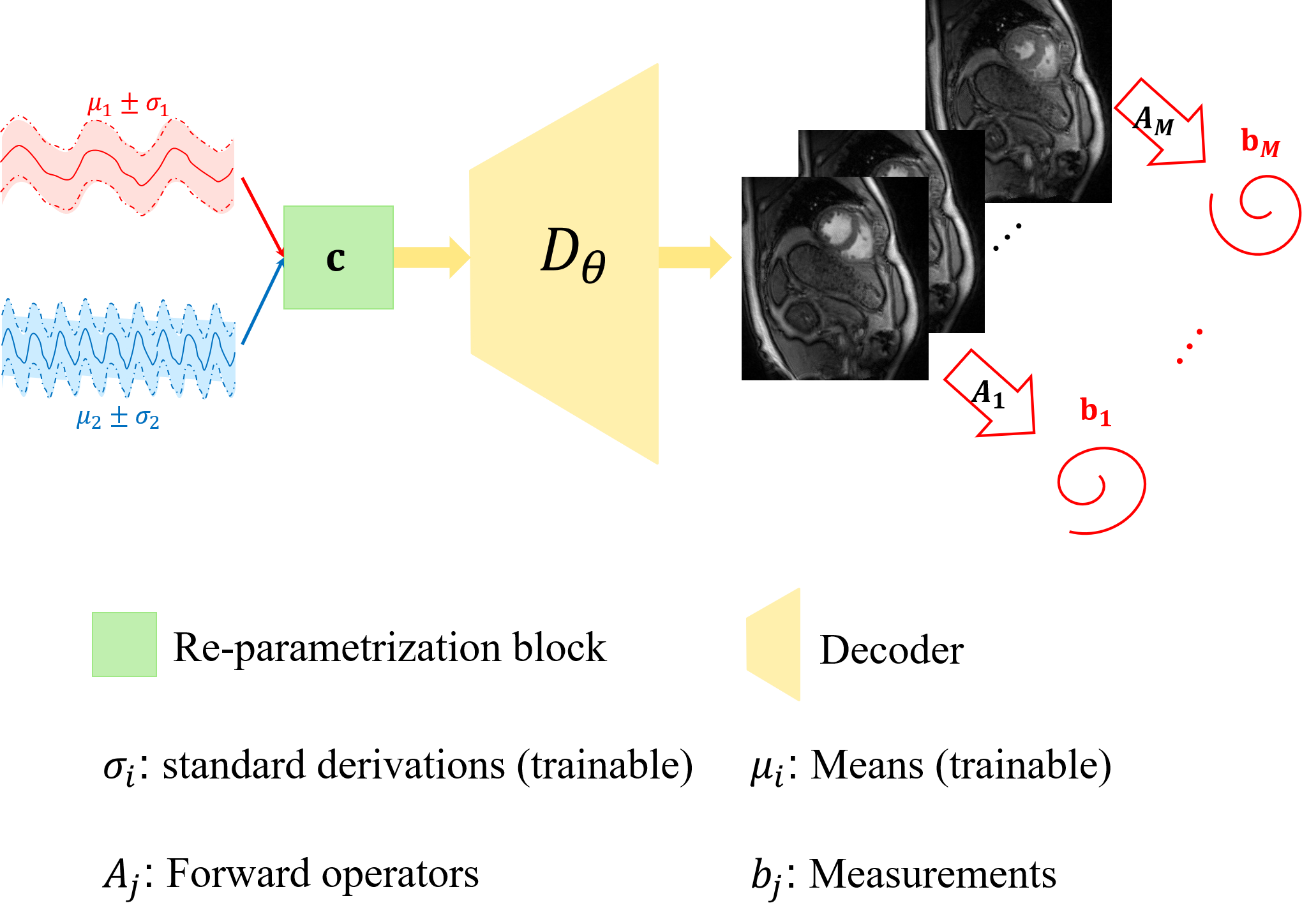}}\hspace{5em}
	\subfigure[V-SToRM: MS]{\includegraphics[width=0.45\textwidth]{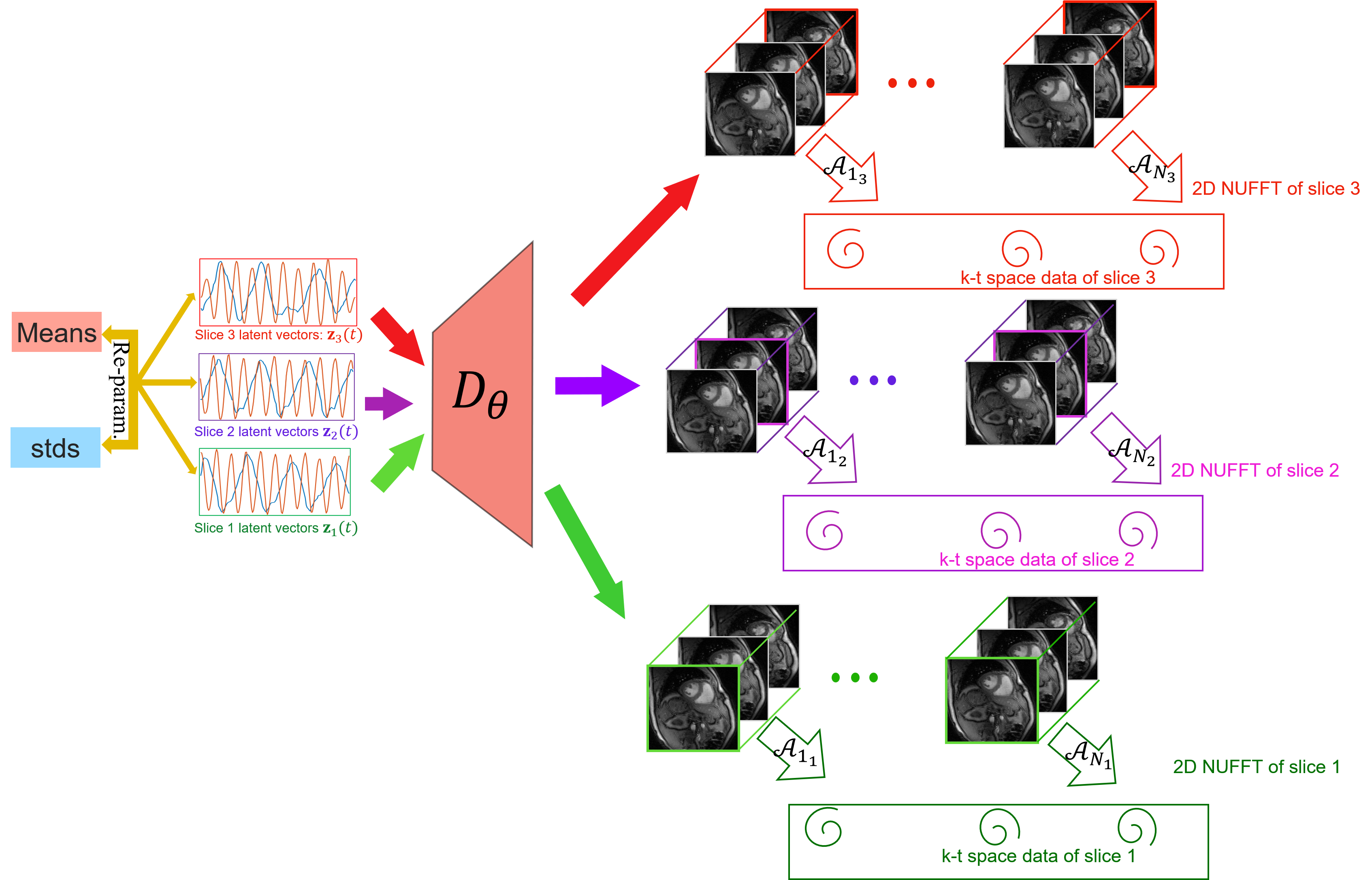}}
	\caption{Illustration of the proposed variational SToRM (V-SToRM) scheme. (a) single-slice setting: The 2D network $\mathcal D$ receives the latent vectors sampled from their respective latent distributions using \eqref{resampling}. The measurements of the 2D-generated images obtained by the respective sampling operators $\mathcal A_i$ are compared to the acquired multi-channel measurements using the cost function specified by \eqref{singleslice}. (b) the multislice 3D setting: Similar to the single-slice setting, the inputs to the 3D network are samples from the respective latent distributions. The 3D volumes are sampled by the respective sampling operators $\mathcal A_{z,t}$, which extract the $z^{\rm th}$ slice and compare it to the measured data. The optimization criterion specified by \eqref{key3} is minimized in this case.}
	\label{illu}
\end{figure*}

\subsection{Single-slice Variational SToRM algorithm}

Based on the analysis in the previous sections, we use the following scheme for the recovery of single-slice dynamic MRI. We use a re-parameterization layer to obtain the latent variables $\mathbf c(t)$ from the time-varying probability distributions $q(\mathbf c(t))$ with parameters  $\boldsymbol\mu_t$ and $\boldsymbol \Sigma_t$. 
These latent variables are fed to the CNN generator $\mathcal D_{\theta}$, which generates the reconstructed volumes $\mathbf x(t) = \mathcal D_{\theta}(\mathbf c(t))$. The multi-channel, non-uniform, Fourier transform-based forward operators are applied on the reconstructed images, which are then compared to the actual noisy measurements $\mathbf{b}_i$. The illustration of this scheme is shown in Fig. \ref{illu} (a). The parameters in the generator and the $\boldsymbol\mu_i$ and the $\boldsymbol \Sigma_i$ are updated based on the loss function 
\begin{equation}\label{singleslice}
\mathcal{L}(\theta,\{\boldsymbol\mu_t, \boldsymbol \Sigma_t\}) = \mathcal C(\theta,\{\boldsymbol\mu_t, \boldsymbol \Sigma_t\}) + \lambda_1||\theta||_1^2 + \lambda_2||\nabla\boldsymbol{\mu}_t||^2.
\end{equation}
Here, $\mathcal C(\theta,\{\boldsymbol\mu_t, \boldsymbol \Sigma_t\})$ is defined in \eqref{vloss}, which is the lower bound for maximum likelihood estimation. The second term in \eqref{singleslice} is a regularization penalty on the generator weights. It has been shown in \cite{zou2021dynamic} that adding this term makes the training of the decoder more stable. The third term involves the temporal gradients of the latent vectors, which enforces the latent vectors to capture the smooth nature of motion patterns in the dynamic images. We use the ADAM optimization to determine the optimal parameters. We also adopt the progressive-in-time training strategy introduced in \cite{zou2021dynamic} to realize a computationally efficient reconstruction. We term this dynamic MRI reconstruction scheme as single-slice variational SToRM.

\subsection{Multislice Variational SToRM algorithm}
\label{multislice}
We now generalize the single-slice variational SToRM scheme for the joint alignment and recovery of multislice dynamic MRI. We assume that the image volume at the time point $t$ during the acquisition of the $z^{\rm th}$ slice, denoted by $\mathbf{x}(\mathbf r,t_z)$, as the output of the generator:
\[\mathbf{x}(\mathbf r,t_z) = \mathcal D_{\theta}\left(\mathbf{c}(t_z)\right).\]
Here, $\mathbf{c}(t_z)$ are the low-dimensional latent vectors corresponding to slice $z$ at the time point $t$, which is formed by the re-parameterization layer. We note that the generator $\mathcal D_{\theta}$ is shared across all slices and time points; this approach facilitates the exploitation of the spatial redundancies between the slices and time points.

We propose to jointly align and reconstruct the multislice MRI by jointly estimating the parameters $\theta$, $\boldsymbol{\mu}(t_z)$ and $\boldsymbol \Sigma(t_z)$ from the measured multislice data by minimizing the following cost function:
\begin{align}\label{key3}\nonumber
\mathcal{L}_{MS}(\theta,\boldsymbol{\mu}(t_z),\boldsymbol \Sigma(t_z)) = & \mathcal C_{MS}(\theta,\boldsymbol{\mu}(t_z),\boldsymbol \Sigma(t_z)) +  \lambda_1||\theta||_1^2\\
& + \lambda_2\sum_z||\nabla_{t_z}\boldsymbol{\mu}(t_z)||^2,
\end{align}
where
\begin{equation*}
\mathcal C_{MS} = \displaystyle\sum_{z=1}^{N_{\rm slice}}\sum_{t=1}^{N_{\rm data}}\|\mathcal{A}_{t_z}\left[\mathcal D_{\theta}(\mathbf c(t_z))\right] - \mathbf b_{t_z}\|^2+\sigma^2~ L(q(t_z))
\end{equation*}
is the lower bound for maximum likelihood as the first term in \eqref{singleslice}. 
The illustration of this scheme is given in Fig. \ref{illu}(b). The parameters of the shared 3D generator $\mathcal D_{\theta}$ are jointly learned in an unsupervised fashion from the measured k-t space data using the ADAM optimization algorithm. After the training process is complete, we will generate the image time series by feeding the generator with the latent variables of any specific slice. Following successful learning, we expect the volumes of the multislice reconstructions to have the same motion patterns characterized by the latent variables of that particular slice. We refer to this dynamic MRI reconstruction scheme as multislice variational SToRM, or V-SToRM.

\subsection{Comparison with state-of-the-art (SOTA) methods}

We compare the proposed V-SToRM approach with the following existing methods.

\begin{itemize}
\item Analysis SToRM \cite{ahmed2020free}: The analysis SToRM model uses a kernel low-rank formulation, which involves the estimation of the manifold Laplacian matrix from the k-space navigators using kernel low-rank regularization. This Laplacian is then used to solve for the images. We note that the analysis SToRM approach has been demonstrated to yield improved performance over state-of-the-art self-gated methods, as shown in our prior work \cite{ahmed2020free,poddar2019manifold}. We refer to this approach as A-SToRM.

\item Single-slice generative SToRM \cite{zou2021dynamic}: The single-slice generative SToRM approach uses a CNN generator to generate the single-slice image series from the highly undersampled k-t space data. This scheme does not rely on a variational formulation. It performs the independent recovery of each slice and hence fails to exploit the inter-slice redundancies. We refer to this approach as G-SToRM:SS.

\item Multislice generative SToRM: We extended the single-slice generative SToRM approach without the variational framework to the multislice setting. In particular, we use the CNN generator to produce the image volume; the generator parameters and the latent vectors for each slice are jointly learned. Finally, we feed the  latent variables of a particular slice into the generator to obtain the aligned multislice reconstruction. We refer to this approach as G-SToRM:MS.
\end{itemize}

For the quantitative comparisons, in addition to the SSIM metric, we also use the signal-to-error ratio (SER) defined as
\[\mathrm{SER} = 20\cdot\log_{10}\frac{||\mathbf{x}_{\rm ref}||}{||\mathbf{x}_{\rm ref}-\mathbf{x}_{\rm recon}||}.\]
Here, $\mathbf{x}_{\rm ref}$ and $\mathbf{x}_{\rm recon}$ represent the reference and the reconstructed images, respectively. The unit for SER is decibel (dB). In our free-breathing and ungated cardiac MRI setting, we usually do not have access to the ground truth. Therefore, in our work, we employ the analysis SToRM method using 25 seconds of data for the reconstruction as the simulated ground truth.

\section{Experiments and results}

\subsection{Implementation details}

In this work, we use deep CNN to build the generator. The number of generator output channels is dependent on the specific datasets. For the experiments using the MNIST dataset, the channel is chosen as 1. By contrast, a two-channel output corresponding to the real and imaginary parts of the MR images is used for the rest of the experiments. In the MRI setting, we use a generator of 10 layers. The total number of trainable parameters is about 6 times the size of the image volume. For the convolutional layers in the generator, the activation function is chosen as leaky ReLU \cite{xu2015empirical} except for the final layer, where $\tanh$ is used as the activation function. Random initialization is used to initialize the generator network. 

The algorithm has three free parameters, $\sigma^2$, $\lambda_1$, and $\lambda_2$. For each method, we optimize these parameters as well as the architecture of the generator on a single dataset such that the reconstructions closely match the 25-second A-SToRM reconstructions. Once the optimal parameters are determined, they are kept fixed for the remaining datasets. Our experiments showed that two latent vectors were sufficient for the good recovery of the single-slice datasets, which correspond to the cardiac and respiratory phases. In the multislice case, we required three to obtain good reconstructions. In this case, two of the three latent vectors captured cardiac and respiratory motion, respectively. The third latent vector seemed to capture a harmonic of the respiratory motion.

\begin{figure}[!htpb]
	\centering
	\includegraphics[width=0.8\textwidth]{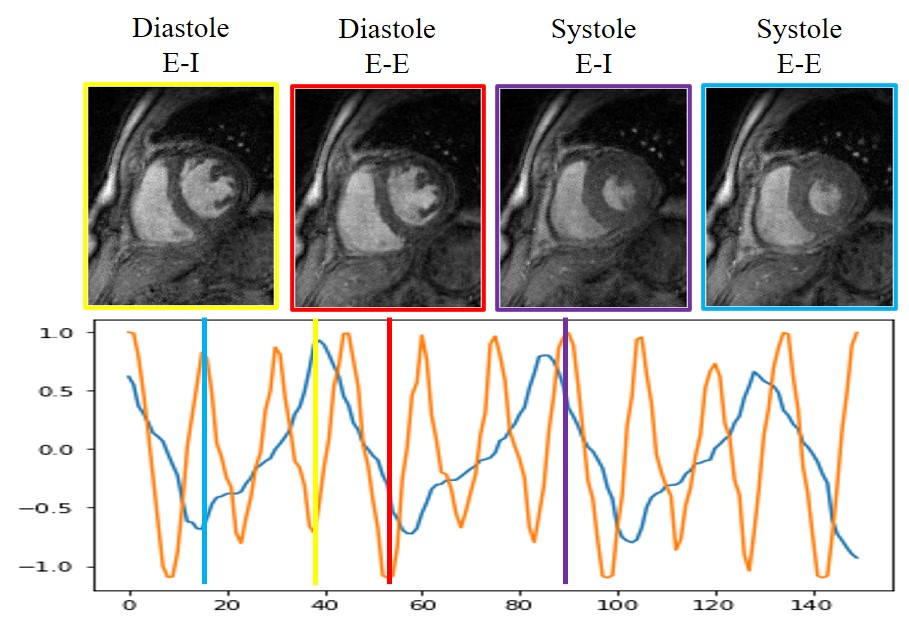}
	\caption{Showcase of the single-slice V-SToRM. We trained the variational model using the data of one slice. We showed four different phases in the time series: diastole in End-Inspiration (E-I), diastole in End-Expiration (E-E), systole in End-Inspiration (E-I), and systole in End-Expiration (E-E), obtained from single-slice V-SToRM. The plot of the latent vectors are shown at the bottom of the figure, and the latent vectors corresponding to the four phases are indicated on the plot of the latent vectors.}
	\label{ss_showcase}
\end{figure}

\subsection{Single-slice V-SToRM and comparisons}

In this section, we focus on single-slice V-SToRM; the reconstructions of a dataset and its latent vectors are shown in Fig. \ref{ss_showcase}. We trained the variational model using the data of one slice. The latent vectors we obtained are shown at the bottom of Fig. \ref{ss_showcase}. Four different phases in the time series are shown in the figure, and their corresponding latent vectors are indicated in the plot of the latent vectors.

\begin{figure}[!htpb]
\centering
	\includegraphics[width=0.8\textwidth]{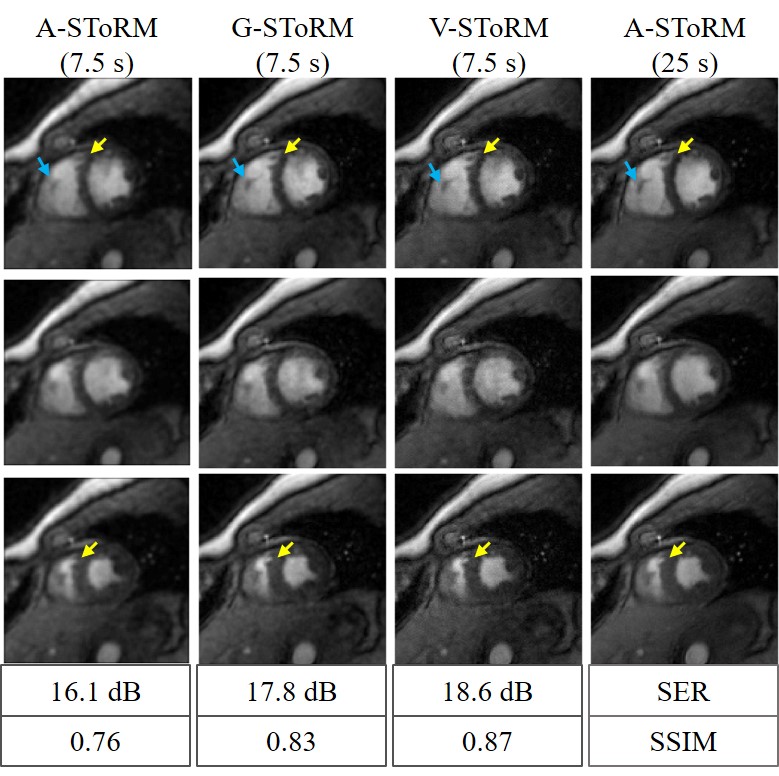}
	\caption{Comparisons with the state-of-the-art methods for single-slice results. The figure shows the visual comparison of three phases: the diastole phase (top row), the systole phase (third row), and the phase that is in between the diastole and systole phases (second row). The first three columns correspond to the reconstructions using the A-SToRM, G-SToRM, and V-SToRM approaches based on 7.5 seconds of data. The last column shows the reconstructions from A-SToRM based on 25 seconds of data; we use these reconstructions as references for quantitative comparisons. We also report the quantitative results at the bottom of the figure.}
	\label{ss_compare}
\end{figure}

The comparisons between the single-slice V-SToRM and the state-of-the-art methods on a different dataset are shown in Fig. \ref{ss_compare}. In these experiments, we compare the region of interest for A-SToRM, G-SToRM, and V-SToRM reconstructions using the 7.5 seconds of data. We use A-SToRM reconstructions from 25 seconds of data as the reference. From Fig. \ref{ss_compare}, we see that G-SToRM (7.5 s) and V-SToRM (7.5) are able to reduce errors and noise in the images when compared to A-SToRM (7.5 s). The proposed V-SToRM (7.5 s) is able to provide sharper edges than G-SToRM (7.5 s). These observations are further confirmed by the quantitative results shown at the bottom of the figure. 

\begin{figure*}[!htpb]
	\centering
	\subfigure[Alignment and recovery of eight slices using V-SToRM]{\includegraphics[width=0.48\textwidth]{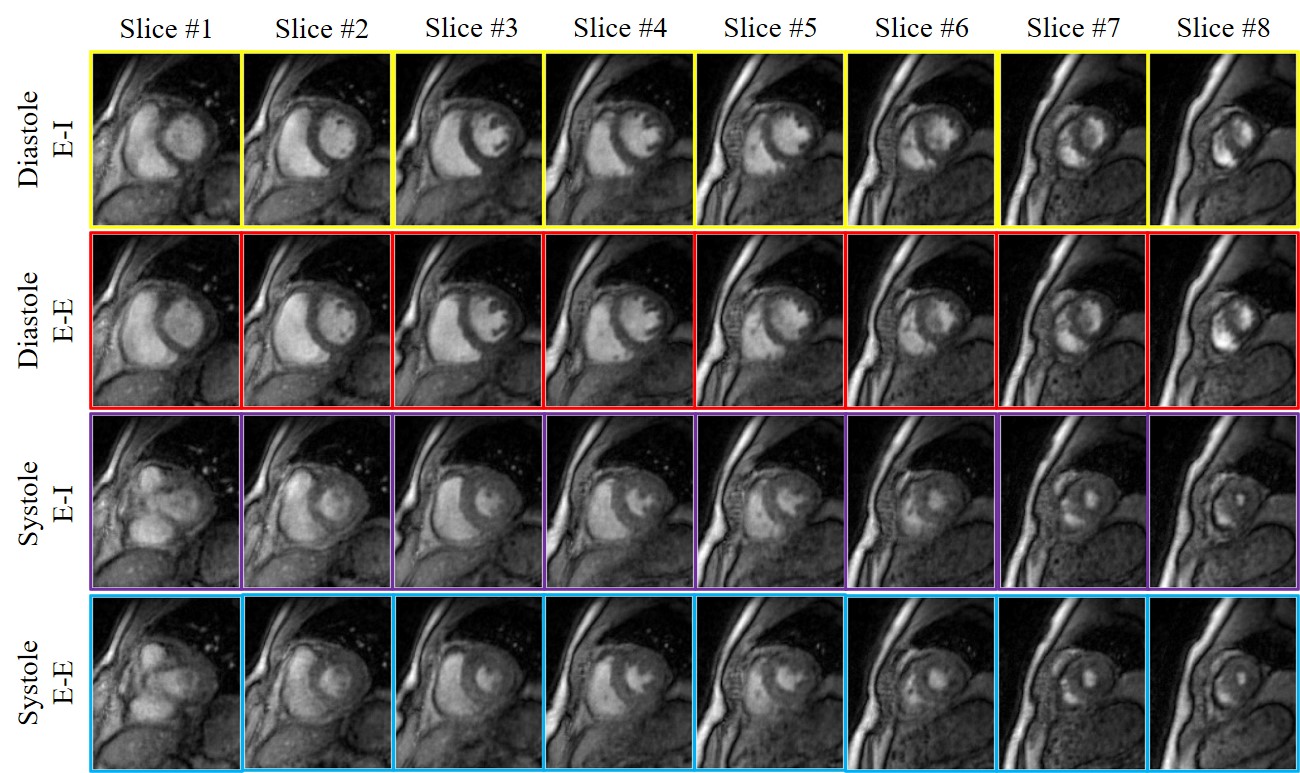}}\hspace{1em}
	\subfigure[Alignment and recovery of eight slices using G-SToRM]{\includegraphics[width=0.48\textwidth]{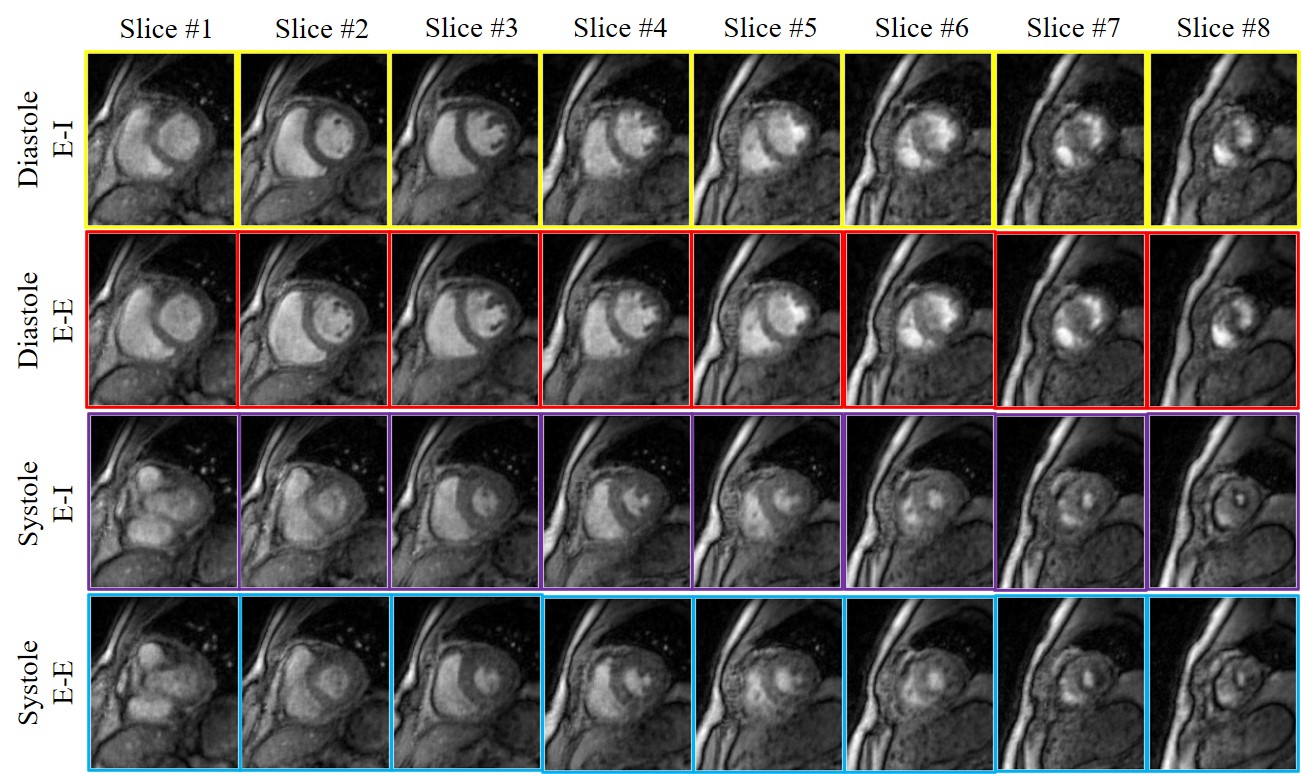}}\\
	\subfigure[Latent vectors obtained by V-SToRM:MS]{\includegraphics[width=0.43\textwidth]{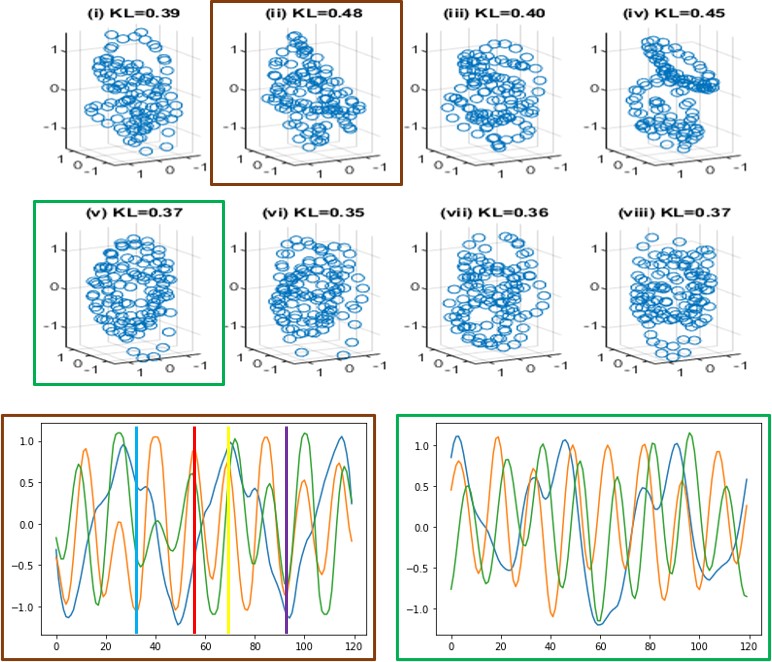}}\hspace{2em}
	\subfigure[Latent vectors obtained by G-SToRM:MS]{\includegraphics[width=0.48\textwidth]{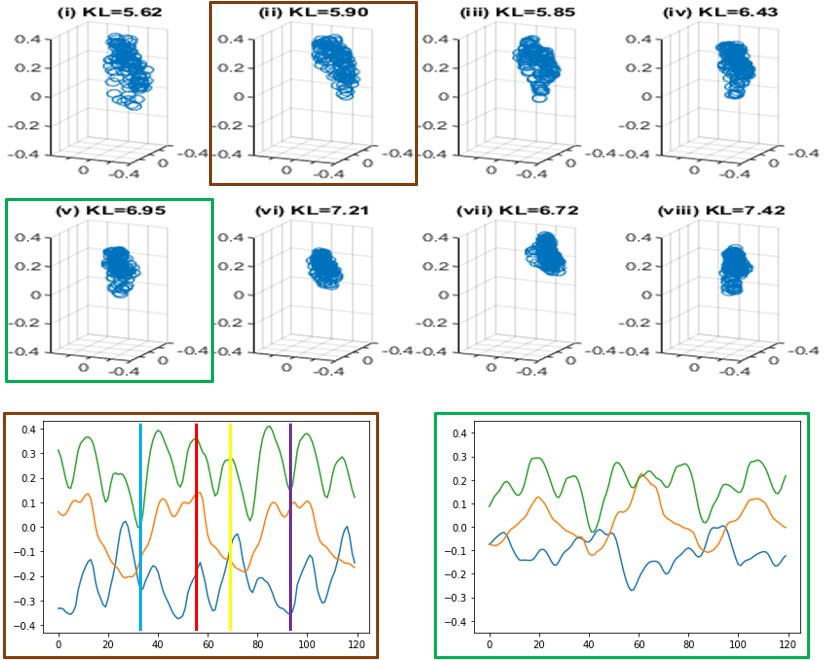}}
	\caption{Alignment and joint recovery of multislice data. In (a), we show the alignment and recovery of the eight slices obtained from the proposed multislice V-SToRM scheme. Four different phases in the time series for each slice are displayed. From (a), we see that all the slices have the same cardiac phase and respiratory phase, indicating that the multislice V-SToRM is able to align the slices. In (b), we show the alignment and recovery of the eight slices obtained from the generalization of single-slice G-SToRM to the multislice setting. We also use four different phases in the time series for each slice to illustrate the alignment of the multislice data. From (b), we see that some of the phases for some of the slices have poor image quality. In particular, the details in the cardiac regions are poorly captured, and in some cases the boundaries of the heart are not visible. These issues can be understood from the plot distributions of the latent vectors obtained by the multislice V-SToRM and G-SToRM:MS, shown in (c) and (d), respectively. We also plot the latent vectors for two of the slices for each method. Note that we generated the results in (a) and (b) by feeding the latent vectors corresponding to the second slice into the generators. The corresponding latent vectors used to generate the four different phases in (a) and (b) are indicated in the plot of the latent vectors in (c) and (d). From (c) and (d), we see that the latent vectors obtained from the proposed multislice V-SToRM scheme have similar distributions, whereas the distributions for the latent vectors obtained from G-SToRM:MS are very different.}
	\label{showcase1}
\end{figure*}

\begin{figure*}[!htpb]
	\centering
	\subfigure[Comparisons based on slice \#3]{\includegraphics[width=0.4\textwidth]{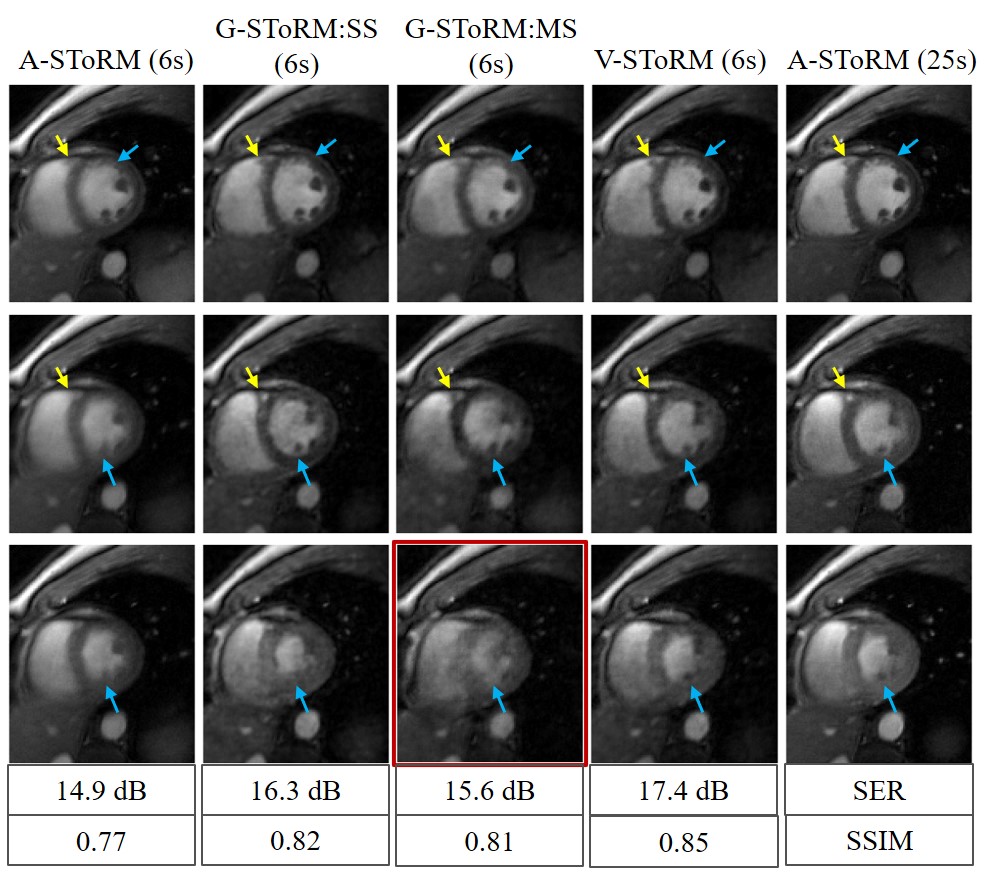}}\hspace{2em}
	\subfigure[Comparisons based on slice \#4]{\includegraphics[width=0.4\textwidth]{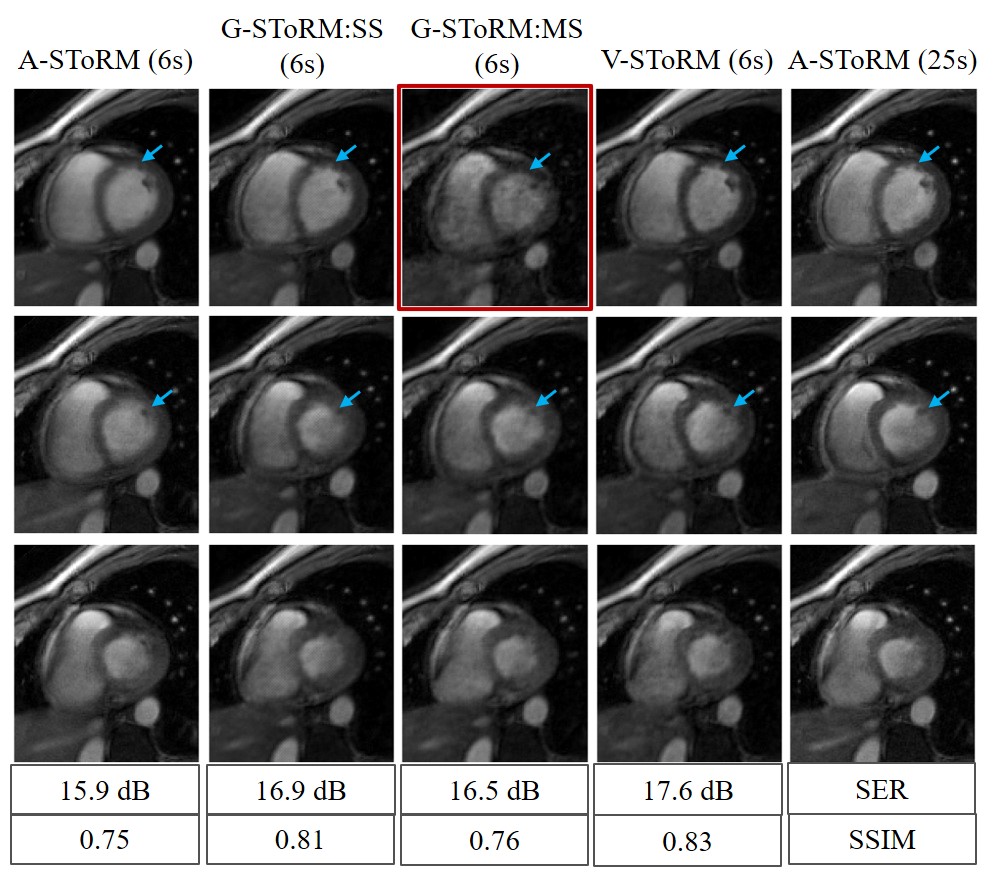}}
	\caption{Comparisons of the image quality of the reconstructions. We compare the image quality of the multislice V-SToRM reconstructions with the image quality of the reconstructions from A-SToRM, G-SToRM:SS, and G-SToRM:MS. The multislice dataset used in this example has four slices, and we show two of the slices in the figure to do the comparisons. For each slice, we show three different phases: the diastole phase, the systole phase, and the phase that is in between the diastole and systole phases. For each sub-figure, the first four columns represent the reconstruction from A-SToRM, G-SToRM:SS, G-SToRM:MS, and the proposed multislice V-SToRM based on 6 seconds of data. The last column shows the reconstructions using A-SToRM based on 25 seconds of data; they are used as simulated references for the quantitative results, which are shown at the bottom of each sub-figure. From both the visual comparisons and the quantitative results, we see that the multislice V-SToRM scheme is able to provide comparable reconstructions when compared to the competing methods. We also highlighted some of the phases in the multislice G-SToRM reconstruction, from which we see that G-SToRM:MS has some issues in generating some of the image frames.}
	\label{ms_compare}
\end{figure*}

\subsection{Joint alignment and recovery of multislice data}

In this section, we show the results of the joint alignment and recovery of multislice data using the proposed multislice V-SToRM scheme. We also compare the alignment results obtained from the straightforward multislice extension of the G-SToRM scheme. The results are shown in Fig. \ref{showcase1}. More results are shown in the supplementary material.

The dataset used in Fig. \ref{showcase1} was acquired with eight slices that covered the whole heart. We trained the variational model based on the undersampled k-t space data and fed the latent vectors corresponding to the second slice to the generator, which produces the aligned multislice reconstructions. Shown in the figures are four time points based on the different phases identified by the latent variables. The rows in Fig. \ref{showcase1} (a) correspond to diastole in End-Inspiration, diastole in End-Expiration, systole in End-Inspiration, and systole in End-Expiration for each slice obtained using the proposed multislice V-SToRM scheme. From Fig. \ref{showcase1} (a), we see that the proposed multislice V-SToRM scheme is able to jointly reconstruct and align the multislice free-breathing and ungated cardiac MRI. We note that all the slices in each row have the same cardiac phase and respiratory phase.

In Fig. \ref{showcase1} (b), we show the corresponding results for the direct extension of the multislice G-SToRM approach. In particular, we trained the model using the undersampled k-t space data and fed the latent vectors corresponding to the second slice into the generator to produce the aligned multislice reconstructions. From Fig. \ref{showcase1} (b), we see that the multislice G-SToRM approach has some ability to align the multislice reconstructions. However, we find that the image quality for some of the frames (e.g., slices 5-8) is poor. For example, the diastole phases for the G-SToRM:MS reconstructions are blurred and the cardiac boundaries are missing. 

The reason for the poor reconstructions offered by multislice G-SToRM and the improved performance of V-SToRM can be easily appreciated from the distribution of the latent vectors shown in Fig. \ref{showcase1} (c) and Fig. \ref{showcase1} (d), respectively. The use of the variational formulation in V-SToRM encouraged the latent variables of the slices to approximate a Gaussian distribution. We also reported the KL divergence value compared to $\mathcal{N}(\mathbf{0},\mathbf{I})$ for each set of the latent vector in the figure. We note that the V-SToRM scheme offers low KL divergence values, indicating that the latent distribution of all the slices are roughly similar to a unit Gaussian. By contrast, the G-SToRM scheme cannot guarantee that the latent variables follow any distribution. We note from the top rows of (d) that the distribution of the latent variables of the second slice is very different from that of the other slices. When we feed the latent vectors of the second slice into the generator, the generator is only able to generate reasonable results for the second slice.

\subsection{Comparison of image quality with state-of-the-art methods}

We compare the image quality of the multislice V-SToRM reconstructions with the image quality of the reconstructions from the state-of-the-art methods, including single-slice methods, in Fig. \ref{ms_compare}. Note that the motion patterns of the slices recovered by the single-slice methods may be very different. For comparison, we manually matched the images of the slices of the single-slice and multislice methods by their cardiac and respiratory phases. The quantitative comparisons of the slices are shown at the bottom of each sub-figure. We also show more results using another dataset in the supplementary material.

The single-slice A-SToRM and G-SToRM:SS comparisons roughly match the observations in Fig. \ref{ss_compare} and the results in \cite{zou2021dynamic}. The results show that the multislice V-SToRM approach is able to offer reconstructions that are less blurred and have fewer alias artifacts when compared to single-slice analysis methods (A-SToRM and G-SToRM:SS). The improved performance is also evidenced by the higher SER and SSIM values. We attribute the improved performance to the exploitation of the redundancies across slices, enabled by V-SToRM. We also note that the G-SToRM:MS method offers poor performance, evidenced by image blurring and missing details on the myocardium. The poor performance of G-SToRM:MS can be understood in terms of the differences in distribution of the latent vectors, shown in Fig. \ref{showcase1}.

\section{Discussion and Conclusion}

In this work, we introduced an approach for the  variational learning of a CNN manifold model from undersampled measurements. This work generalized the traditional VAE scheme to the undersampled setting. Unlike the traditional VAE scheme that uses an encoder to learn the conditional distribution from the images, we propose to learn the parameters of the distribution from the measurements using back-propagation. The application of the framework to multislice cardiac MRI data enabled the joint alignment and recovery  from highly undersampled measurements. Unlike current single-slice methods that perform independent recovery of the slices, the proposed approach aligns the acquisitions and jointly recovers the images from the undersampled k-t space data. In addition to facilitating the exploitation of inter-slice redundancies, this approach also eliminates the need for post-processing schemes to match the phases of the slices. 

Our results show that the joint alignment and recovery of the slices offer reduced blurring and reduction of artifacts compared to the direct generalization of G-SToRM to the multislice setting. In particular, the variational framework encourages the latent variables of different slices to have the same distribution. By contrast, the G-SToRM framework cannot guarantee the similarity of the probability distributions; the improper alignment translates to image blurring and other artifacts. Similarly, the use of the CNN generator offers implicit spatial regularization, resulting in improved recovery over A-SToRM. 

A benefit with the proposed scheme is that it does not require fully sampled data to train the CNN. The subject-specific CNN parameters and the latent vectors are learned directly from the undersampled data. We note that the acquisition of fully sampled data to train neural networks is not always possible, especially in the high-resolution and dynamic MRI settings considered in this work. In this context, direct learning from undersampled data is desirable. However, a challenge of the proposed scheme when compared to pretrained deep learning methods that offer super-fast inference is the higher computational complexity. We will explore training strategies, including transfer learning and meta-learning, to reduce the run time in the future.

\section{Appendix}

In this appendix, we show that the likelihood term in \eqref{likelihood} can be lower-bounded by \eqref{elbo}.

According to \eqref{likelihood} and using the result of joint probability, we obtain 
\begin{eqnarray}\nonumber\label{probab1}
p(\mathbf{b}_i) &=& \frac{p(\mathbf{b}_i,\mathbf{c}_i)}{q_i(\mathbf{c}_i)} \frac{q_i(\mathbf{c}_i)}{p(\mathbf{c}_i|\mathbf{b}_i)}\\
&=& \underbrace{\frac{p(\mathbf{b}_i,\mathbf{c}_i)}{p(\mathbf{c}_i)}}_{p(\mathbf{b}_i|\mathbf{c}_i)}\frac{p(\mathbf{c}_i)}{q_i(\mathbf c_i)} \frac{q_{i}(\mathbf c_i)}{p(\mathbf c_i|\mathbf b_i)}.
\end{eqnarray}
Taking the logarithm on both sides of \eqref{probab1}, we have
\begin{equation}\label{probab2}
\log p(\mathbf{b}_i) = \log p(\mathbf{b}_i|\mathbf{c}_i)-\log\frac{q_i(\mathbf c_i)}{p(\mathbf{c}_i)}+\log \frac{q_i(\mathbf c_i)}{p(\mathbf c_i|\mathbf b_i)}.
\end{equation}
Next, we take the expectation with respect to $\mathbf{c}_i\thicksim q_i(\mathbf c_i)$ of both sides of \eqref{probab2}, and realizing that $\mathop{\mathbb{E}}_{\mathbf{c}_i\thicksim q_i(\mathbf c_i)}\log p(\mathbf{b}_i) = \log p(\mathbf{b}_i)$, we obtain
\begin{eqnarray}\nonumber\label{probab3}
\displaystyle \log p(\mathbf{b}_i) =& \displaystyle\underbrace{\displaystyle \mathop{\mathbb{E}}_{\mathbf{c}_i\thicksim q_i(\mathbf c_i)} \log p(\mathbf{b}_i|\mathbf{c}_i)}_{{\text{data term}}} 
- \underbrace{\displaystyle \mathop{\mathbb{E}}_{\mathbf{c}_i\thicksim q_i(\mathbf c_i)}\log\frac{q_i(\mathbf c_i)}{p(\mathbf{c}_i)}}_{KL[q_i(\mathbf c_i)||p(\mathbf{c}_i)] } \\
&+ \underbrace{\displaystyle \mathop{\mathbb{E}}_{\mathbf{c}_i\thicksim q_i(\mathbf c_i)} \log \frac{q_i(\mathbf c_i)}{p(\mathbf c_i|\mathbf b_i)}}_{KL[q_i(\mathbf c_i)||p(\mathbf{c}_i|\mathbf{b}_i)] >0}.
\end{eqnarray}
The last term is always greater than zero. The first term is the conditional density of the measurements $\mathbf{b}_i$ given the images $\mathbf{x}_i= \mathcal D_{\theta}(\mathbf c_i)$. With the measurement model specified by \eqref{measurements}, we obtain
\[\displaystyle \mathop{\mathbb{E}}_{\mathbf{c}_i\thicksim q_i(\mathbf c_i)}\log~p(\mathbf{b}_i|\mathbf{c}_i) = -\frac{1}{2\sigma^2} \displaystyle \mathop{\mathbb{E}}_{\mathbf{c}_i\thicksim q_i(\mathbf c_i)}\|\mathcal{A}_i \,\mathcal D(\mathbf{c}_i)-\mathbf b_i\|^2 + c,\]
where $c$ is a constant independent of the parameters of interest. Ignoring the constant $c$ and plugging $\mathbb{E}_{\mathbf{c}_i\thicksim q_i(\mathbf c_i)}\log~p(\mathbf{b}_i|\mathbf{c}_i)$ back into \eqref{probab3}, we obtain the desired lower bound \eqref{elbo}.


\section*{Acknowledgments}

The authors would like to thank Ms. Melanie Laverman from the University of Iowa for making editorial corrections to refine this paper. Financial support for this study was provided by grants NIH 1R01EB019961 and NIH R01AG067078-01A1. This work was conducted on MRI instruments funded by 1S10OD025025-01.

\bibliographystyle{IEEEbib}
\bibliography{refs}

\end{document}